\documentclass[12pt]{article}
\usepackage{amsfonts,amssymb,epsfig,psfig,cite}

\begin{document}


\def\reff#1{(\ref{#1})}
\newcommand{\be}{\begin{equation}}
\newcommand{\ee}{\end{equation}}
\newcommand{\<}{\langle}
\renewcommand{\>}{\rangle}

\def\spose#1{\hbox to 0pt{#1\hss}}
\def\ltapprox{\mathrel{\spose{\lower 3pt\hbox{$\mathchar"218$}}
 \raise 2.0pt\hbox{$\mathchar"13C$}}}
\def\gtapprox{\mathrel{\spose{\lower 3pt\hbox{$\mathchar"218$}}
 \raise 2.0pt\hbox{$\mathchar"13E$}}}

\def\bsigma{\mbox{\protect\boldmath $\sigma$}}
\def\bpi{\mbox{\protect\boldmath $\pi$}}
\def\smfrac#1#2{{\textstyle\frac{#1}{#2}}}
\def\smhalf{ {\smfrac{1}{2}} }

\newcommand{\re}{\mathop{\rm Re}\nolimits}
\newcommand{\im}{\mathop{\rm Im}\nolimits}
\newcommand{\tr}{\mathop{\rm tr}\nolimits}
\newcommand{\fr}{\frac}
\newcommand{\diti}{\frac{\mathrm{d}^2t}{(2 \pi)^2}}

\def\Z{{\mathbb Z}}
\def\R{{\mathbb R}}
\def\C{{\mathbb C}}

\title{Renormalised four-point coupling constant \\ in the three-dimensional 
$O(N)$ model with $N\to0$}

\author{ Andrea Pelissetto  \\[0.5mm]
  Dipartimento di Fisica and INFN -- Sezione di Roma I  \\
  Universit\`a degli Studi di Roma ``La Sapienza'' \\
  Piazzale Moro 2, I-00185 Roma, Italy \\
  e-mail: {\rm Andrea.Pelissetto@roma1.infn.it } \\ [1.5mm]
  Ettore Vicari  \\[0.5mm]
  Dipartimento di Fisica and INFN -- Sezione di Pisa  \\
  Universit\`a degli Studi di Pisa \\
  Largo Pontecorvo 2, I-56127 Pisa, Italy \\
  e-mail: {\rm Ettore.Vicari@df.unipi.it }}

\maketitle
\thispagestyle{empty}   

\begin{abstract}
  
  We simulate self-avoiding walks on a cubic lattice and
  determine the second virial coefficient for walks of different lengths. This
  allows us to determine the critical value of the renormalized four-point
  coupling constant in the three-dimensional $N$-vector universality class for
  $N=0$.  We obtain $\bar{g}^* = 1.4005(5)$, where $\bar{g}$ is normalized so
  that the three-dimensional field-theoretical $\beta$-function behaves as
  $\beta(\bar{g}) = - \bar{g} + \bar{g}^2$ for small $\bar{g}$.  As a
  byproduct, we also obtain precise estimates of the interpenetration ratio
  $\Psi^*$, $\Psi^* = 0.24685(11)$, and of the exponent $\nu$, $\nu =
  0.5876(2)$.



\end{abstract}

\clearpage

\section{Introduction}

In the last thirty years there has been significant progress in the
understanding of critical phenomena. The conceptual setting is now well
understood and we are now in a position to check the general framework by
comparing experimental and theoretical results and the different theoretical
methods among themselves. The most precise experimental and theoretical
results have been obtained for $O(N)$ systems in which the order parameter is
an $N$-component vector and the symmetry breaking pattern corresponds to
$O(N)\to O(N-1)$ \cite{PV-review}. With the increase of the precision of
theoretical and experimental estimates, some small discrepancies are beginning
to emerge: for instance, there is at present a discrepancy between the
experimental \cite{LSNCI-96,LNSSC-03} and the theoretical
\cite{CHPRV-01,BMPS-06,CHPV-06} estimates of the specific-heat exponent
$\alpha$ for the three-dimensional XY universality class ($N=2$); analogously,
there are tiny discrepancies between the most precise field-theoretical
estimates \cite{GZ-98} of $\gamma$ for three-dimensional Ising ($N=1$) and
polymer systems ($N=0$) and those obtained by using high-temperature and Monte
Carlo methods \cite{CPRV-02,DB-03,CCP-98,HNG-04}.  These differences 
should not be taken as an indication of the failure of the general
framework; most likely, they are due to a too optimistic determination of the
error bars on the results. For instance, numerical results are affected by
scaling corrections which are difficult to take into account, while
field-theoretical results may converge slowly due to non-analyticities of the
renormalization-group functions
\cite{Nickel-82,Nickel-91,Sokal-94,PV-98,CPV-01,CCCPV-00}.\footnote{
  Non-analyticity effects are expected to be small in three dimensions.  For
  instance, the leading non-analytic terms appearing in the $\beta$ function
  have the form $(g^* - g)^{1+1/\Delta}$ and $(g^* - g)^{\Delta_2/\Delta}$
  \cite{Nickel-91,PV-98}.  In three dimensions, $1+1/\Delta\approx
  \Delta_2/\Delta \approx 2$ so that non-analyticities are weak
  \cite{PV-review,CPV-01}. This is not the case in two dimensions
  \cite{CCCPV-00}. In the two-dimensional Ising model, a correction of the
  form $(g^* - g)^{8/7}$ is expected.  This term is probably the reason why
  field-theory results \cite{OS-00} differ significantly from the exact ones
  \cite{CPV-01,CCCPV-00}.  
}

The case $N=0$ is a good testing ground for the different theoretical methods.
Indeed, there is a well known mapping between the $N=0$ universality class and
the statistical model of self-avoiding walks (SAWs) \cite{deGennes-79}.  SAWs
can be very efficiently simulated by using several algorithms.  In particular,
in three dimensions the pivot algorithm
\cite{Lal-69,MJHS-85,MS-88,Sokal-95b,Kennedy-02} allows one to obtain an
independent measurement of a global quantity in a time of the order of
$\xi^{0.85/\nu} \sim \xi^{1.45}$ (in SAWs the correlation length $\xi$ can be
identified with the end-to-end distance) \cite{Kennedy-02}, to be compared
with conventional algorithms for spin systems in which the autocorrelation
time scales as $\xi^{3+z}$, where $z$ is the dynamic critical exponent
(cluster algorithms, which are at present the best ones for $N$-vector models,
have $0 < z \lesssim 0.5$; see \cite{CB-92,OS-04} for estimates of $z$ and
additional references). Thus, for $N=0$ one is able to probe very carefully
the critical limit and get a good control of the scaling corrections that
represent the main source of error in high-precision studies. Monte Carlo
studies provide therefore accurate estimates of critical quantities that can
be used as reference values in other theoretical approaches.

In this paper we consider the four-point renormalized coupling which is the
basic ingredient in any field-theoretical calculation, computing its value at
the critical point. We obtain 
\begin{equation}
\bar{g}^{*} = 1.4005(5),
\label{eq1}
\end{equation}
where $\bar{g}$ is normalized so that the $\beta$ function behaves as
$\beta(\bar{g}) = - \bar{g} + \bar{g}^2 + O(\bar{g}^3)$ for small values of
$\bar{g}$. This result should be compared with the best available estimates.
Field theory gives $\bar{g}^{*} = 1.396(20)$ ($\epsilon$ expansion
\cite{PV-00}) and $\bar{g}^* = 1.413(6)$ (massive zero-momentum scheme in
fixed dimension $d = 3$ \cite{GZ-98}).  The $\epsilon$-expansion result is
perfectly consistent with our final estimate. One the other hand, the
fixed-dimension result differs slightly, by approximately two error bars. This
is similar to what is observed for the exponent $\gamma$: the most precise
fixed-dimension calculation gives \cite{GZ-98} $\gamma = 1.1596(20)$, which is
slightly larger---but fully compatible with the quoted error---than the most
precise Monte Carlo estimates, $\gamma = 1.1573(2)$ \cite{HNG-04} and $\gamma
= 1.1575(6)$ \cite{CCP-98}. A similar phenomenon occurs for $N=1$, where the
fixed-dimension estimates \cite{GZ-98} of both $\bar{g}^{*}$ and $\gamma$ are
slightly larger (but note that again the effect is at the level of one error
bar) than the Monte Carlo results \cite{CPRV-02,DB-03}.  The universal
constant $\bar{g}^*$ has also been computed by resumming its high-temperature
expansion in the lattice $N$-vector model.  Ref.~\cite{BC-98} reports
$\bar{g}^* = 1.388(5)$, which is not fully compatible with (\ref{eq1}).  As
stressed several times \cite{PV-98,PV-review}, this is probably due to scaling
corrections proportional to $(\beta - \beta_c)^\Delta$, where $\Delta$ is the
leading correction-to-scaling exponent ($\Delta \approx 0.5$ in the present
model). Even if in principle standard resummation methods should be able to
take them into account, in practice, with the series of moderate length
available today, they give rise to systematic deviations that are quite
difficult to estimate: as a consequence, error bars are often underestimated.
This problem has been overcome by considering improved Hamiltonians
characterized by the absence of leading scaling corrections
~\cite{CPRV-02,CHPV-06,CHPRV-02}, leading to accurate estimates
of critical exponents and universal amplitudes for $N=1,2,3$.

In Table \ref{tablegstar} we summarize the best available estimates of
$\bar{g}^{*}$ for three-dimensional $N$-vector systems with $0\le N \le 3$.
Monte Carlo or high-temperature expansions computed in improved models provide
the most accurate results. Field theory is in substantial agreement, although
small differences appear for $N=0$ (fixed-dimension expansion) and $N=3$
($\epsilon$ expansion).  Other results are reported in \cite{PV-review}.

\begin{table}
\begin{center}
\begin{tabular}{ccllll}
\hline\hline
 Reference & Method & $N=0$ & $N=1$ & $N=2$ & $N=3$ \\
\hline
\cite{GZ-98} & FD & 1.413(6) & 1.411(4) & 1.403(3) & 1.390(4) \\
\cite{PV-00} & $\epsilon$-exp & 
                    1.396(20) & 1.408(13) & 1.425(24) & 1.426(9) \\
\cite{CPRV-02,CHPV-06,CHPRV-02} & IHT & & 1.406(1) & 1.4032(7) & 1.395(7) \\
present work & MC   & 1.4005(5) & & & \\
\hline\hline
\end{tabular}
\end{center}
\caption{Estimates of the critical value of the four-point
renormalized coupling $\bar{g}^*$ in the three-dimensional 
$N$-vector model with $0\le N \le 3$. 
We normalize $\bar{g}$ so that 
the three-dimensional $\beta$ function behaves 
as $\beta(\bar{g}) = - \bar{g} + \bar{g}^2 + O(\bar{g}^3)$. In 
\cite{CPRV-02,CHPV-06,CHPRV-02} a different normalization is used and 
$g^* = 48 \pi \bar{g}^*/(N+8)$ is reported. In the second column we indicate the 
method used in the determination of the coupling constant. FD refers to the 
field-theoretical 
fixed-dimension zero-momentum scheme, $\epsilon$-exp to the $\epsilon$ expansion, 
MC stands for Monte Carlo and IHT for the analysis of high-temperature expansions
specialized to improved models (i.e. models without leading scaling corrections).}
\label{tablegstar}
\end{table}

The paper is organized as follows.  In Sec.~\ref{sec2} we define the basic
quantities that are computed in the present work. In Sec.~\ref{sec3} we
present our numerical results. First we derive the correction-to-scaling
function associated with the second virial coefficient, then we determine
$\bar{g}^*$. Finally, in Sec.~\ref{sec3.3} we use our numerical data to obtain
a new estimate of the exponent $\nu$.

\section{Definitions} \label{sec2}

We consider a simple cubic lattice and $N$-step SAWs. A SAW is a
lattice walk $\{{\bf r}_0,\cdots,{\bf r}_N\}$ such that ${\bf r}_i$ and ${\bf
  r}_{i+1}$ are nearest neighbors and each lattice site is visited at most
once.  We also introduce an effective attraction $-{\cal E}$ (${\cal E} > 0$)
between nonconnected nearest-neighbor walk sites. If $\beta \equiv {-\cal
  E}/k_B T$ is the dimensionless inverse temperature, in the scaling limit the
statistical properties are independent of $\beta$ as long as the system is in
the good-solvent (swollen) regime. Scaling corrections instead depend on
$\beta$ and thus, by properly fixing $\beta$, one can reduce them
significantly.  Following \cite{PH-05}, we fixed $\beta = 0.054$. The
extensive Monte Carlo work of \cite{CMP-06-vir} indicates that for this value
of $\beta$ the leading scaling corrections are at least a factor-of-10 smaller
than those occurring in the athermal model with $\beta =
0$.\footnote{According to \cite{CMP-06-vir} scaling corrections vanish for
  $\beta = 0.048(7)$. 
Our simulations started before reference
  \cite{CMP-06-vir} was completed.  For this reason we used the older estimate
  of \cite{PH-05}.}  For the simulation, we used the pivot algorithm
\cite{Lal-69,MJHS-85,MS-88,Sokal-95b,Kennedy-02}, which is very efficient for
the determination of SAW global properties.  In order to estimate the
zero-momentum renormalized coupling $g^*$, we
computed the second virial coefficient for SAWs of different lengths
\cite{MN-87}.  For this purpose we used the hit-or-miss algorithm discussed in
\cite{LMS-95}.

\begin{table}
\begin{center}
\begin{tabular}{crrl}
\hline\hline
$\lambda$ & $N_1$ & $N_2$ & $V_2(N_1,N_2;\beta)$ \\
\hline
0.1250 &  8000 & 64000 & 0.38972(20)\\ 
0.1406 &  9000 & 64000 & 0.38564(20)\\ 
0.1563 & 10000 & 64000 & 0.38249(20)\\ 
0.2135 & 10250 & 48000 & 0.37362(20)\\ 
0.2240 & 10750 & 48000 & 0.37236(18)\\ 
0.2292 & 11000 & 48000 & 0.37154(19)\\ 
0.2656 &  8500 & 32000 & 0.36787(19)\\ 
0.2969 &  9500 & 32000 & 0.36561(17)\\ 
0.4063 &  9750 & 24000 & 0.35963(14)\\ 
0.4432 &  9750 & 22000 & 0.35821(13)\\ 
0.4625 &  9250 & 20000 & 0.35757(8)\\ 
0.5000 & 12000 & 24000 & 0.35670(15)\\ 
0.5455 & 12000 & 22000 & 0.35536(13)\\ 
0.5469 &  8750 & 16000 & 0.35523(17)\\ 
0.7188 & 11500 & 16000 & 0.35320(15)\\ 
0.7609 &  8750 & 11500 & 0.35286(15)\\ 
0.8000 &  8000 & 10000 & 0.35278(15)\\ 
0.8125 &  9750 & 12000 & 0.35268(15)\\ 
0.8889 &  8000 &  9000 & 0.35235(14)\\ 
0.8947 &  8500 &  9500 & 0.35217(15)\\ 
0.9000 &  9000 & 10000 & 0.35248(14)\\ 
0.9167 & 22000 & 24000 & 0.35210(13)\\ 
0.9318 & 10250 & 11000 & 0.35243(12)\\ 
0.9535 & 10250 & 10750 & 0.35241(15)\\ 
0.9697 &  8000 &  8250 & 0.35214(13)\\ 
0.9773 & 10750 & 11000 & 0.35241(15)\\ 
\hline\hline
\end{tabular}
\end{center}
\caption{Estimates of $V_2(N_1,N_2;\beta)$ for $\beta = 0.054$.
We report the Monte Carlo results such that $N_2>N_1\ge 8000$.}
\label{table-V}
\end{table}

We performed a large-scale simulation, considering walks of length $N$
varying between 100 and 64000, determining the radius of gyration and the 
end-to-end distance for 67 different values of $N$ and the 
second virial coefficient for 369 pairs of walks of different length. 
Some results (those corresponding to pairs with $N_2> N_1\ge 8000$)
are reported in Table \ref{table-V}.
The statistics
vary between $8\times 10^7$ and $24\times 10^7$ pivot trials for each value of $N$.
Our runs were performed on a cluster of Intel Xeon (3.20 GHz) processors
and lasted approximately 8 CPU years of a single processor.

We measured the following quantities:
\begin{itemize}
\item[(1)] The radius of gyration of a SAW of length $N$,
\be
R^2_g(N;\beta) \equiv  
{1\over 2 (N+1)^2} \sum_{i,j} \< (\mathbf{r}_i - \mathbf{r}_j)^2 \>,
\ee
where the sums go over the $N+1$ sites of the chain and $\mathbf{r}_i$ 
is the corresponding position.
\item[(2)] The end-to-end distance of a SAW of length $N$,
\be
R^2_e(N;\beta) \equiv  
\< (\mathbf{r}_0 - \mathbf{r}_N)^2 \>.
\ee
\item[(3)] The second virial coefficient for two SAWs of length 
$N_1$ and $N_2$,
\be
B_2(N_1,N_2;\beta) \equiv  {1\over2} \sum_{\mathbf{r}}\,
     \< 1 - e^{-\beta H(1,2)}\>_{\mathbf{0},\mathbf{r}},
\label{B2def}
\ee
where the average is over two walks of length $N_1$ and $N_2$, the first
one starting at the origin and  the second at $\mathbf{r}$.
The sum is over all lattice sites and
$H(1,2)$ is interaction energy between the two chains:
$H(1,2)=+\infty$ if the two walks intersect each other; otherwise,
$H(1,2)=- {\cal E} N_{nnc}$, where $N_{nnc}$ is the number of lattice bonds
$\langle \mathbf{r}_a \mathbf{r}_b\rangle$, such that $\mathbf{r}_a$ belongs to the 
the first walk and $\mathbf{r}_b$ belongs to the second walk (or vice versa).
\end{itemize}
The second virial coefficient is not universal. A universal
quantity is obtained by considering the following 
dimensionless ratio:
\begin{eqnarray}
V_2 (N_1,N_2;\beta) &\equiv & {B_2(N_1,N_2;\beta)\over 
       R_e(N_1;\beta)^{3/2} R_e(N_2;\beta)^{3/2}} .
\end{eqnarray}
In the scaling limit, i.e. for $N_1,N_2\to \infty$, we have
\begin{eqnarray}
R^2(N_1;\beta) &=& a_R(\beta) N_1^{2\nu} (1 + b_R(\beta) N_1^{-\Delta} + \cdots) \\
V_2 (N_1,N_2;\beta) &=& V^*(\lambda) + 
    b_V(\beta) f(\lambda) (N_1 N_2)^{-\Delta/2} + \cdots 
\label{expV2}
\end{eqnarray}
where $\lambda \equiv  N_1/N_2$, $f(1) = 1$ (normalization condition) and 
we have neglected additional subleading scaling corrections. The 
functions $V^*(\lambda)$ and $f(\lambda)$ as well as the 
exponents $\nu$ and $\Delta$ are universal. On the other hand,
the amplitudes $a_R(\beta)$, $b_R(\beta)$ and $b_V(\beta)$ are model-dependent
and therefore depend explicitly on $\beta$.\footnote{The amplitude ratio 
$b_R(\beta)/b_V(\beta)$ is universal and therefore $\beta$ independent. 
Estimates are reported in \cite{CMP-06-rad}: $b_{R_g}/b_V = -1.5(3)$,
$b_{R_e}/b_V = -1.1(3)$. Also the ratio $a_{R_g}/a_{R_e}$ is universal.
Reference \cite{CMP-06-rad} reports $A_{ge}^* \equiv (a_{R_g}/a_{R_e})^2 = 0.15988(4)$.} 
The exponent $\nu$ is known quite precisely. 
At present the most accurate estimates are
$\nu = 0.58758 \pm 0.00007$ \cite{BN-97},
$\nu = 0.5874 \pm 0.0002$ \cite{Prellberg-01} and
$\nu = 0.58765 \pm 0.00020$ \cite{HNG-04}
(for an extensive list of results, see \cite{PV-review}).
In Sec.~\ref{sec3.3} we confirm these results, obtaining 
$\nu = 0.5876 \pm 0.0002$. Also the exponent $\Delta$
has been determined quite accurately
\cite{BN-97}: $\Delta = 0.515\pm 0.007^{+0.010}_{-0.000}$.
In Ref.~\cite{CMP-06-vir} the second virial coefficient for walks 
of equal length was determined, leading to the estimate 
$V^*(1) (A_{ge}^*)^{-3/2} = 5.500(3)$. Using \cite{CMP-06-rad} 
$A_{ge}^* \equiv (a_{R_g}/a_{R_e})^2 = 0.15988(4)$, we obtain 
$V^*(1) = 0.3516(2)$.

Given $V^*(\lambda)$, the critical value of the four-point renormalized coupling constant
is given by \cite{MN-87}
\be
\bar{g}^* = {6^{3/2}\over \pi} 
   {\Gamma(3\nu + 2\gamma)\over \Gamma(\gamma)^{1/2} \Gamma(\gamma + 2\nu)^{3/2}} 
\int_0^\infty 
d\lambda \;   \lambda^{3\nu/2 + \gamma - 1} (1 + \lambda)^{-3\nu - 2\gamma} 
   V^{*}(\lambda)\;.
\label{gstar-lambda}
\ee
We use here the standard field-theoretical 
normalization in which the $O(N)$ $\beta$ function behaves 
as $\beta(\bar{g}) = - \bar{g} + \bar{g}^2$ for small $\bar{g}$. 
In (\ref{gstar-lambda}) the universal 
critical exponent $\gamma$ appears. At present the most precise estimates are
$\gamma = 1.1575(6)$ \cite{CCP-98} and $\gamma = 1.1573(2)$ \cite{HNG-04}.

Let us note that $V_2(N_1,N_2;\beta)$ is symmetric under the interchange of $N_1$ and 
$N_2$. This implies that $V^*(\lambda)$ and $f(\lambda)$ are both symmetric
under the transformation $\lambda \to 1/\lambda$. In order to make this 
symmetry explicit we introduce a new variable
\be
\mu \equiv  {2\lambda\over 1 + \lambda^2},
\ee
which varies in the interval [0,1] and is symmetric under $\lambda \to 1/\lambda$.
We take $\mu$ as fundamental variable, considering $V^*$ and $f$ to be 
functions of the variable $\mu$. In terms of $\mu$, integral (\ref{gstar-lambda})
becomes
\be
\bar{g}^* = {6^{3/2}\over \pi} 
   {\Gamma(3\nu + 2\gamma)\over \Gamma(\gamma)^{1/2} \Gamma(\gamma + 2\nu)^{3/2}} 
\int_0^1 
   {2 d\mu\over \mu \sqrt{1 - \mu^2}}
     \left[{2 (1 + \mu)\over \mu} \right]^{-3\nu/2 - \gamma} V^{*}(\mu)\; .
\label{gstar-mu}
\ee
We wish now to compute the small-$\mu$ behavior of the scaling functions.
For this purpose we extend a scaling argument due to de Gennes 
\cite{deGennes-82}. For $N_1/N_2\to 0$, we can compute the second virial 
coefficient by dividing the longest walk (of length $N_2$) in $N_2/N_1$ blobs, which have 
a size of the order of the size of the shortest walk (of length $N_1$). 
The shortest SAW interacts only with a single blob, so that 
$
B(N_1,N_2) \sim (N_2/N_1) B_{\rm bl}(N_1),
$
where $B_{\rm bl}(N_1)$ is the second virial coefficient that takes into account the 
interaction between the blob and the shortest walk.  Since $B_{\rm bl}(N_1)$ 
depends on a single length scale, the length $N_1$, by dimensional reasons 
we have
$B_{\rm bl}(N_1)\sim N_1^{3\nu} (1 + c N_1^{-\Delta})$. 
It follows
\be
B(N_1,N_2) \sim N_2 N_1^{3\nu-1} (1  + c N_1^{-\Delta}),
\ee
and then
\be
V_2(N_1,N_2) \sim {N_2 N_1^{3\nu-1} (1  + c N_1^{-\Delta}) \over 
      N_1^{3\nu/2} (1 + d_1 N_1^{-\Delta}) 
      N_2^{3\nu/2} (1 + d_2 N_2^{-\Delta})  } \sim 
      \lambda^{3\nu/2-1} [1 + k \lambda^{-\Delta/2} (N_1 N_2)^{-\Delta/2}]\; .
\ee
Comparing with the expansion (\ref{expV2}) and taking into account that, for small
$\lambda$, $\lambda\approx \mu/2$, we obtain
\begin{equation}
V^*(\mu)\sim \mu^{3\nu/2-1}, \qquad
f(\mu)\sim \mu^{3\nu/2-1-\Delta/2}.
\label{smallmu-beh}
\end{equation}
As a consequence of this result, for $\mu\to 0$ the integrand
that appears in (\ref{gstar-mu}) behaves as 
$\mu^{3\nu+\gamma-2} \sim \mu^{0.92}$. Thus, the small-$\mu$
region is not crucial for a precise determination of $\bar{g}^*$.
This is very welcome, since it is quite difficult to determine 
$V^*(\mu)$ precisely for $\mu$ small.

\section{Numerical results}\label{sec3}

\subsection{Determination of the correction-to-scaling function $f(\mu)$}
\label{sec3.1}

\begin{figure}
\centerline{\epsfig{file=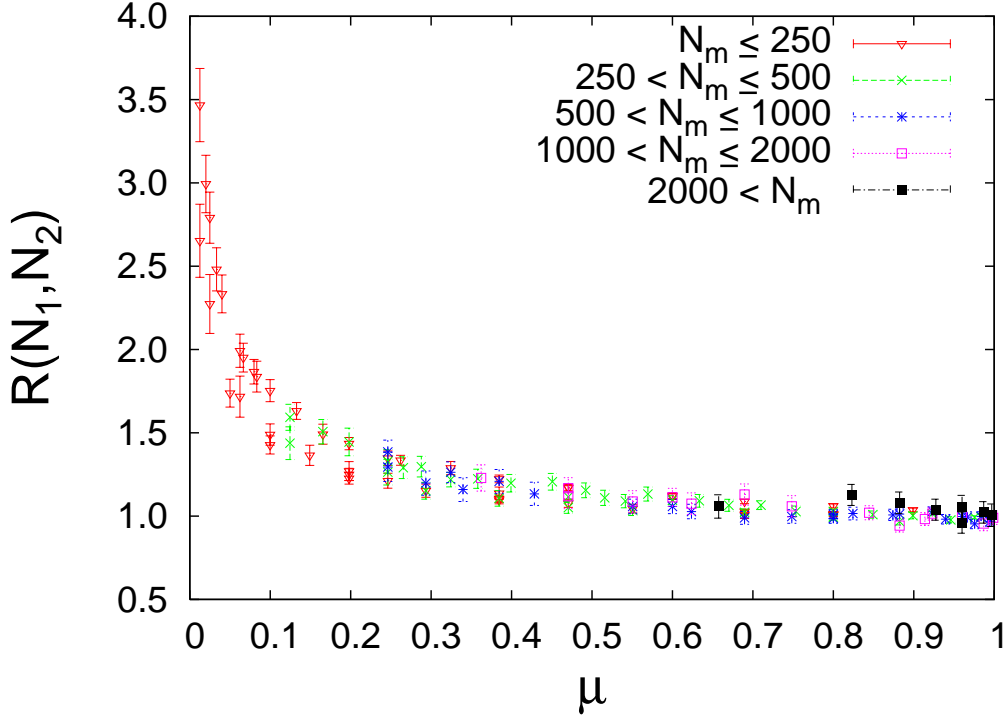,angle=-90,width=14truecm}}
\caption{Plot of $R(N_1,N_2)$ as a function of 
$\mu \equiv 2 N_1 N_2/(N_1^2 + N_2^2)$, for $\Delta = 0.515$. 
We use different symbols according to the 
value of $N_m = \min(N_1,N_2)$.}
\label{figR}
\end{figure}

In order to determine the correction-to-scaling function $f(\mu)$ 
[see (\ref{expV2})], 
we use the numerical data of \cite{PH-05,Pelissetto-06}.
The function $V_2(N_1,N_2;\beta)$
was determined for several values of $N_1$,$N_2$ in the range 
$50\le N_1,N_2 \le 16000$ for $\beta = 0$ and $\beta = 0.1$,
though not so precisely as in the present work.
Following \cite{PH-05} we consider 
\be
R(N_1,N_2) \equiv  \left({N_1\over N_2}\right)^{-\Delta/2} 
  {V_2(N_1,N_2;\beta = 0) - V_2(N_1,N_2;\beta = 0.1) \over 
   V_2(N_1,N_1;\beta = 0) - V_2(N_1,N_1;\beta = 0.1) },
\label{defR}
\ee
which converges to $f(\mu)$ as $N_1,N_2\to \infty$. In Fig.~\ref{figR}
we show $R(N_1,N_2)$ versus $\mu$. The data approximately fall
onto a single curve. However, at a closer look, one observes that,
especially for small values of $\mu$, the data fall 
on two different, though close, lines. This fact can be easily
understood. 
The function $R(N_1,N_2)$ defined in (\ref{defR}) 
is not symmetric under the interchange of $N_1$
and $N_2$ and thus, for finite values of $N_1$
and $N_2$, there is no symmetry under $\lambda\to1/\lambda$. 
This symmetry is recovered only in the scaling limit.
Thus, given a value of $\mu$, the data cluster around two different
values: one corresponds to a value of $\lambda$ such that $\lambda > 1$,
while the second one corresponds to a value of $\lambda$ such that $\lambda < 1$.
These differences disappear in the scaling limit and thus they provide us  
an estimate of the next-to-leading scaling corrections.

It is possible to define a quantity which is symmetric under the interchange
of $N_1$ and $N_2$ and converges to $f(\mu)$ in the scaling limit.
We could have defined 
\be
R'(N_1,N_2) \equiv  
  {V_2(N_1,N_2;\beta = 0) - V_2(N_1,N_2;\beta = 0.1) \over 
   [\Delta V_2(N_1) \Delta V_2(N_2)]^{1/2} },
\ee
where $\Delta V_2(N) \equiv  V_2(N,N;\beta = 0) - V_2(N,N;\beta = 0.1)$. 
We could not use this definition since estimates of $\Delta V_2(N)$ are 
available only for a few values of $N$ (those computed in \cite{PH-05}), 
and thus $R'(N_1,N_2)$ can be determined only for 16 pairs of walks.
         
\begin{figure}
\centerline{\epsfig{file=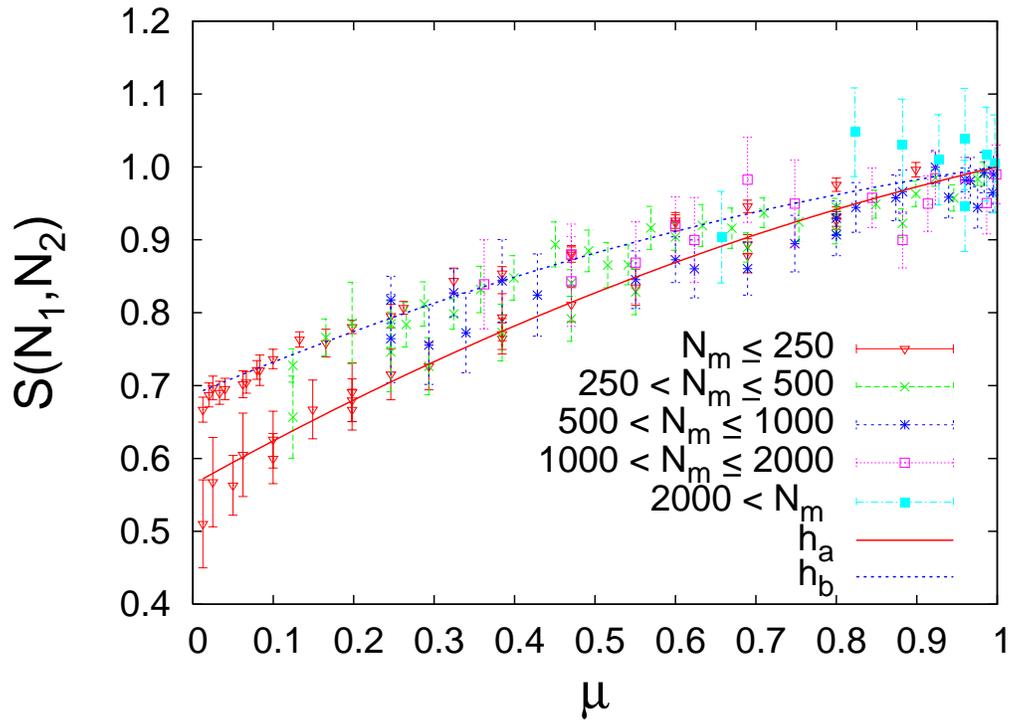,angle=-90,width=14truecm}}
\caption{Plot of $S(N_1,N_2)$ as a function of 
$\mu \equiv  2 N_1 N_2/(N_1^2 + N_2^2)$, for 
$\Delta = 0.515$ and $\nu = 0.58758$. We use different symbols according to the 
value of $N_m = \min(N_1,N_2)$. We also report the functions $h_a(\mu)$ and 
$h_b(\mu)$ discussed in the text, see (\ref{hestimate}).}
\label{figS}
\end{figure}

Since the function $f(\mu)$ behaves as $\mu^{3\nu/2 - 1 - \Delta/2}$ as 
$\mu\to 0$ [equation (\ref{smallmu-beh})], we write 
\be
f(\mu) = \mu^{3\nu/2 - 1 - \Delta/2} h(\mu),
\ee
where $h(\mu)$ satisfies $h(0) \not = 0$ and the normalization
condition $h(1) = 1$. In order to estimate $h(\mu)$ we consider
\be
S(N_1,N_2) \equiv  R(N_1,N_2)
   \left( {2 N_1 N_2\over N^2_1 + N_2^2} \right)^{-3\nu/2 + 1 + \Delta/2},
\ee
which converges to $h(\mu)$ in the scaling limit. The function $S(N_1,N_2)$ is 
reported in Fig.~\ref{figS}. As before, note that the data do not fall
onto a single curve: two branches, corresponding respectively to 
$\lambda > 1$ and $\lambda<1$ are clearly visible. Their difference gives us a rough
estimate of the next-to-leading corrections. The two branches of the function 
$S(N_1,N_2)$ are quite smooth in $\mu$ and thus good fits
are obtained by taking  polynomial interpolations.
Therefore, we fit the numerical data to
\be
S(N_1,N_2) = 1 + \sum_{k=1}^n a_k (\mu -1)^k,
\ee 
which automatically guarantees $S(N,N) = 1$. In order to take into 
account the scaling corrections that show up in the presence of two
different branches, we perform two fits: in the first one 
[fit (a)] we use all data for $\mu\ge 0.7$ and the data with $N_1 > N_2$ 
for $\mu< 0.7$; in the second one [fit (b)] 
we use all data for $\mu\ge 0.7$ and the data with $N_1 < N_2$
for $\mu< 0.7$. In case (a) we interpolate the data that
belong to the lower branch, while in case (b) we interpolate
the upper-branch results. The order $n$ of the polynomial 
does not play much role and we always use $n=2$. The results of the 
two fits are
\begin{eqnarray}
h_a(\mu) &=& 1 + 0.2559953 (\mu - 1) - 0.179692 (\mu - 1)^2, \nonumber \\
h_b(\mu) &=& 1 + 0.1604295 (\mu - 1) - 0.152394 (\mu - 1)^2.
\label{hestimate}
\end{eqnarray}
The corresponding curves
are reported in Fig.~\ref{figS}. In the following section we use
\be
h(\mu) = {1\over2}[h_a(\mu) + h_b(\mu)] = 
         1 + 0.208212 (\mu - 1) -  0.166043 (\mu - 1)^2
\label{hestimate2}
\ee
as our estimate of $h(\mu)$ and take $|h - h_a|$ and $|h - h_b|$
as estimates of the error on $h(\mu)$.

\subsection{Estimate of $g^*$} \label{sec3.2}

\begin{figure}
\centerline{\epsfig{file=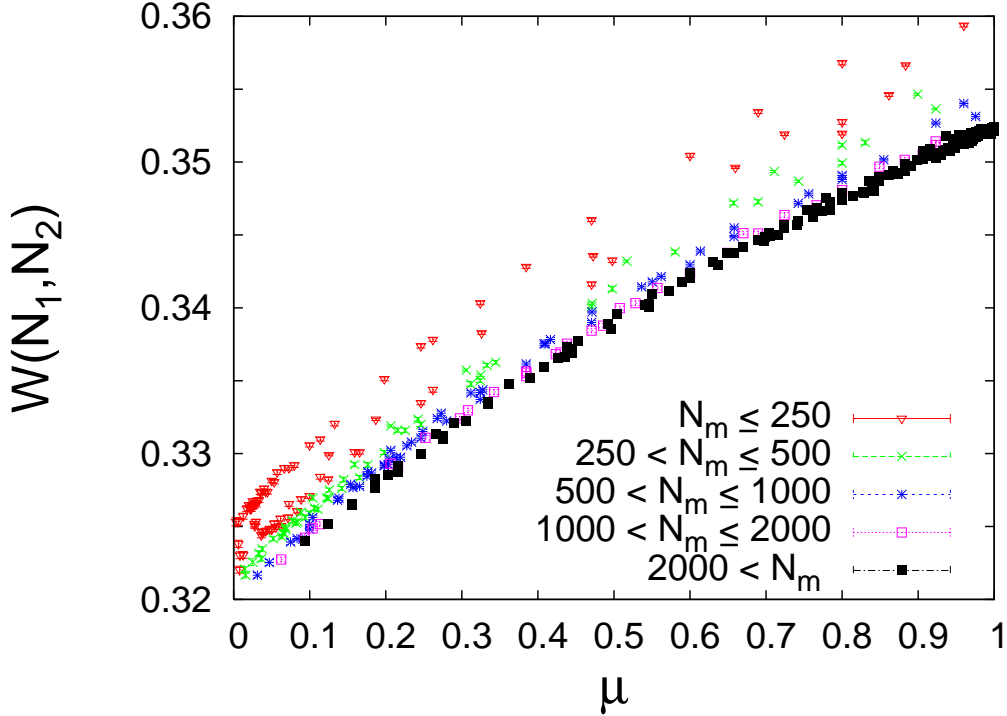,angle=-90,width=14truecm}}
\caption{Plot of $W(N_1,N_2;\beta)$ for $\beta = 0.054$ as a function of 
$\mu \equiv  2 N_1 N_2/(N_1^2 + N_2^2)$ for $\nu = 0.58758$. 
We use different symbols according to the 
value of $N_m = \min(N_1,N_2)$. }
\label{figW}
\end{figure}

We now estimate the universal function $V^*(\mu)$. Taking into account
the asymptotic behavior (\ref{smallmu-beh}), we define a new function
\be
W(\mu) \equiv  \mu^{1-3\nu/2} V^*(\mu),
\ee
which is such that $W(0) \not= 0$. In order to estimate $W(\mu)$ we 
consider
\be
W(N_1,N_2;\beta) = V_2(N_1,N_2;\beta)
\left( {2 N_1 N_2\over N^2_1 + N_2^2} \right)^{-3\nu/2 + 1}.
\ee
For $N_1,N_2\to \infty$ it behaves as [see (\ref{expV2})]
\be
W(\mu) + b_V(\beta) h(\mu) \mu^{-\Delta/2} (N_1 N_2)^{-\Delta/2}.
\ee
The function $W(N_1,N_2;\beta)$ is reported in Fig.~\ref{figW}. 
It shows a small dependence on $\mu$, but significant scaling corrections.
As before we use a polynomial parametrization for $W(\mu)$ and fit the data to
\be
W(N_1,N_2;\beta) = \sum_{k=0}^n a_k (\mu-1)^k + 
          a_{n+1} h(\mu) \mu^{-\Delta/2} (N_1 N_2)^{-\Delta/2},
\label{param-W}
\ee
where we use the function $h(\mu)$ reported in (\ref{hestimate2}),
and $a_0,\cdots, a_{n+1}$ are fit parameters. 
Once $W(\mu) = \sum_{k=0}^n a_k (\mu-1)^k$ has been determined, 
we can compute $V^*(\mu)$ and 
finally $\bar{g}^*$ using (\ref{gstar-mu}).
 There are several sources of error on the result:
\begin{itemize}
\item[i)] Statistical errors due to the uncertainty on $V_2(N_1,N_2;\beta)$. They are
computed by means of an auxiliary Monte Carlo procedure. The input data are 
varied randomly within error bars (at each Monte Carlo step 
all estimates $V_2(N_1,N_2)$ are replaced by 
$V_2(N_1,N_2) + r(N_1,N_2) \sigma_V(N_1,N_2)$, where $\sigma_V(N_1,N_2)$ is the 
error on $V_2(N_1,N_2)$ and $r(N_1,N_2)$ is a random number extracted from a 
Gaussian distribution with zero mean and unit variance) and $\bar{g}^*$ 
is recomputed 
each time. The standard deviation of these estimates provides the statistical 
error on $\bar{g}^*$.
\item[ii)] Error due to the uncertainty on $h(\mu)$. 
We repeat the analysis with the 
two functions reported in (\ref{hestimate}).
The difference with the result obtained
by using $h(\mu)$ gives the error.
\item[iii)] Error due to the uncertainty of the exponents. We consider the 
best estimates $\nu = 0.58758(7)$, $\Delta = 0.515(17)$, and $\gamma = 1.1573(2)$,
and determine how $\bar{g}^*$ varies as the exponents change by one error bar.
\item[iv)] Error due to the additional scaling corrections.
In (\ref{expV2}) we have only written the leading scaling correction. 
There are however other correction terms that vanish faster and 
that may be relevant 
at the values of $N$ we are considering. For this purpose we have 
repeated the analysis several times, each time considering only data 
satisfying $N_1,N_2\ge N_{\rm min}$, for several values of $N_{\rm min}$.
\end{itemize}

\begin{table}
\begin{center}
\begin{tabular}{rcc}
\hline\hline
$N_{\rm min}$ & $\bar{g}^*$ & $V^*(1)$ \\
\hline
100 &  1.3983(5) & 0.35108(16)\\
200 &  1.3990(5) & 0.35120(14)\\
400 &  1.3997(4) & 0.35134(11)\\
800 &  1.4003(4) & 0.35147(10)\\
1000 &  1.4004(4) & 0.35151(10)\\
1500 &  1.4005(4) & 0.35153(10)\\
2000 &  1.4004(5) & 0.35150(10)\\
3000 &  1.4005(5) & 0.35149(12)\\
5000 &  1.4007(7) & 0.35154(19)\\
\hline\hline
\end{tabular}
\end{center}
\caption{Estimates of $\bar{g}^*$ and of $V^*(1)$ for different
values of $N_{\rm min}$. A third-order interpolation ($n=3$) is used 
in (\ref{param-W}).}
\label{tablegV}
\end{table}

In Table \ref{tablegV} we give estimates of $\bar{g}^*$ 
for different values of $N_{\rm min}$. We use a third-order interpolation,
i.e. we set $n=3$ in (\ref{param-W}). Similar results are obtained for $n=2$: for 
$N_{\rm min} \lesssim 2000$ the estimates obtained by taking $n=2$ and $n=3$
differ by less than ${1\over3}$ of the error bar; for $N_{\rm min} \gtrsim 2000$ 
differences are larger bur never exceed one error bar. The error we report
is the sum of the errors of type (i), (ii) and (iii). Errors of type (ii)
are always small and account for less than 5\% of the total error: the somewhat 
large uncertainty on the function $h(\mu)$ does not have much influence 
on the final result. For $N_{\rm min} \lesssim 1000$ most of the error
is due to the error on the exponent $\Delta$, while in the opposite
case the statistical error (i) dominates. As a check of our results we also report
$V^*(1)$. With the parametrization (\ref{param-W}) we have simply 
$V^*(1) = a_0$. The results are perfectly consistent with the estimate
$V^*(1) = 0.3516(2)$ reported in Sec.~\ref{sec2}, obtained by using the 
numerical results of \cite{CMP-06-vir,CMP-06-rad}. 

The results reported in Table \ref{tablegV} show a tiny dependence on 
$N_{\rm min}$. For $N_{\rm min}\ge 800$ 
the results are approximately constant within error bars.
We take our final estimates at $N_{\rm min} = 3000$:
\begin{eqnarray}
\bar{g}^* &=& 1.4005(5), \label{estg} \\
V^*(1) &=&  0.35149(12). 
\label{estV1}
\end{eqnarray}
Estimate (\ref{estg}) should be compared with previous results:
$\bar{g}^* = 1.396(20)$ obtained by using the $\epsilon$ expansion
\cite{PV-00}; $\bar{g}^* = 1.413(6)$ obtained by using the massive
zero-momentum scheme in three dimensions \cite{GZ-98}; $\bar{g}^* =
1.388(5)$ obtained by resumming its high-temperature expansion \cite{BC-98}.
Note that both the fixed-dimension estimate and the high-temperature result
are not fully consistent with our results. In the case of the
field-theoretical results this is probably due to the slow convergence of the
perturbative expansions, a phenomenon that may be related to the
non-analyticities of the renormalization-group functions at the fixed point.
In the case of the high-temperature result this is probably due to the scaling
corrections: resummations are not fully able to cope with the
non-analyticities present at the critical point.

Our precise estimate of $\bar{g}^*$ can also be used as input in 
field-theory determinations of the critical exponents $\gamma$ and $\nu$. 
We consider the massive zero-momentum scheme 
\cite{GZ-98} in fixed dimension,
in which critical quantities are obtained 
by evaluating the corresponding renormalization-group functions at 
$\bar{g}^*$. 
Instead of using the six-loop estimate $\bar{g}^*=1.413(6)$, we
evaluate the seven-loop resummed renormalization-group functions associated
with $\gamma$ and $\nu$
\cite{BNGM-77,MN-91,GZ-98} at the Monte Carlo estimate
$\bar{g}^*=1.4005(5)$.  This can be easily achieved by using some intermediate
results reported in \cite{GZ-98}.  We obtain
\begin{equation}
\gamma=1.1583(15),\qquad \nu=0.5873(7),
\label{newft}
\end{equation}
which are slightly lower than the estimates of \cite{GZ-98}, 
$\gamma=1.1596(20)$ and $\nu=0.5882(11)$, and in better
agreement with the best Monte Carlo estimates
$\gamma=1.1573(2)$ and $\nu=0.58758(7)$ mentioned above.\footnote{One may repeat this
  calculation for other values of $N$. For example, in the Ising case
  ($N=1$), using the most precise estimate $\bar{g}^*=1.406(1)$, see Table
  \ref{tablegstar}, one obtains $\gamma=1.2387(8)$ and $\nu=0.6299(10)$, which
  should be compared with the field-theoretical estimates \cite{GZ-98}
  $\gamma=1.2396(13)$ and $\nu=0.6304(13)$ (obtained by using the six-loop result
  $\bar{g}^*=1.411(4)$), and the lattice results 
  \cite{CHPRV-02,DB-03} $\gamma=1.2373(2)$ and $\nu=0.63012(16)$.}

The estimate (\ref{estV1}) of $V^*(1)$ allows us to obtain a new estimate of the 
interpenetration ratio:
\be
\Psi^* \equiv  2 (4\pi)^{-3/2} (A_{ge}^*)^{-3/2} V^*(1) = 0.24685(11),
\ee
where we used $A_{ge}^* = 0.15988(4)$ \cite{CMP-06-rad}. It is consistent 
with the result reported in \cite{CMP-06-vir}: 
$\Psi^* = 0.24693(13)$.

\subsection{Estimate of the exponent $\nu$} \label{sec3.3}

\begin{table}
\begin{center}
\begin{tabular}{rll}
\hline\hline
$N_{\rm min}$ &   $\nu(R_e^2)$ & $\nu(R_g^2)$ \\
\hline
100 & 0.58768(9)  & 0.58754(10) \\
200 & 0.58773(14) & 0.58755(14) \\
300 & 0.58761(18) & 0.58754(17) \\
400 & 0.58756(19) & 0.58751(19) \\
500 & 0.58761(23) & 0.58753(23) \\
600 & 0.58750(25) & 0.58743(24) \\
800 & 0.58764(33) & 0.58768(31) \\
1000 & 0.58740(36) & 0.58741(34) \\
\hline\hline
\end{tabular}
\end{center}
\caption{Estimates of $\nu$ obtained from the analysis of 
$R_e^2$ and $R_g^2$ as a function of $N_{\rm min}$, 
the length of the shortest walk considered in the fit.}
\label{tablenu}
\end{table}

As a byproduct of our simulations we obtained estimates 
of $R^2_g$ and $R^2_e$ up to $N=64000$. They allow us to 
obtain a new estimate of the exponent $\nu$. It is crucial to 
take into account the scaling corrections and thus we have performed 
fits of the form
\be
\ln R^2 = a + 2\nu\ln N + b N^{-\Delta} + c N^{-\Delta_2},
\label{fitR2}
\ee
as we did in \cite{CMP-06-vir,CMP-06-rad} for the virial 
coefficients, taking $\Delta = 0.515(17)$ \cite{BN-97} and 
$\Delta_2 = 1.0(1)$. The last term in (\ref{fitR2})
is an effective correction that takes into account several terms:
nonanalytic corrections proportional to $N^{-2\Delta}$ and 
$N^{-\Delta_2}$ ($\Delta_2$ is the next-to-leading correction-to-scaling
exponent, $\Delta_2 = 0.98(6)$ \cite{NR-84}) and analytic terms 
behaving as $1/N$. As before, we repeat the fit several times,
each time including data corresponding to walks such that 
$N\ge N_{\rm min}$. The results are reported in Table~\ref{tablenu}.
The estimates obtained from the fits of $R_g^2$ are very stable---they 
essentially do not change for $100\le L_{\rm min} \le 500$. Those obtained from
the fits of $R_e^2$ show a slight downward trend but are perfectly compatible.
As final estimate we take
\be
\nu = 0.5876(2),
\label{estnu}
\ee
which is compatible with all results reported in Table~\ref{tablenu}. 
The estimate (\ref{estnu}) is in good agreement with the previous
ones reported in the literature \cite{BN-97,Prellberg-01,HNG-04} 
and mentioned in Section \ref{sec2}. 

\begin{figure}
\centerline{\epsfig{file=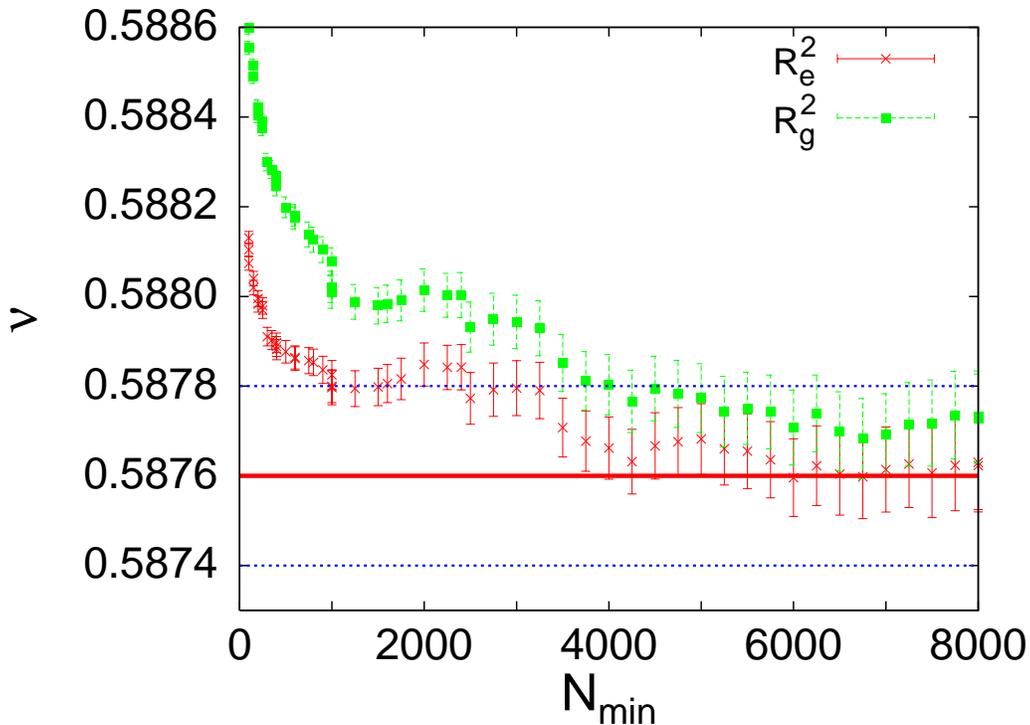,angle=-90,width=14truecm}}
\caption{Estimates of $\nu$ obtained by fitting $R^2_g$ and $R^2_e$ to
$\ln R^2 = a + 2\nu \ln N$, as a function of $N_{\rm min}$, 
the length of the shortest walk considered in the fit. The thick horizontal line
corresponds to our central estimate (\ref{estnu}), while the thin dotted lines give the
error.}
\label{fignu}
\end{figure}

Note that inclusion of the scaling corrections in the fit is crucial.
In Fig.~\ref{fignu} we report the estimates of $\nu$ obtained by using 
a fit without scaling corrections, $\ln R^2 = a + 2\nu\ln N$.
The results show a clear downward trend, depend on the quantity at hand, and 
are apparently constant within error bars only for $N_{\rm min} \gtrsim 6000$.
In this range they are in perfect agreement with the estimate (\ref{estnu}).

\subsection*{Acknowledgements}

The numerical results presented here have been obtained 
on the Theory cluster at CNAF (INFN) in Bologna.

\end{document}